\documentstyle[12pt,epsf,epsfig,wrapfig]{article}
\begin{document}

\begin{center}
{\bfseries CHARMONIUM POLARIZATION IN HIGH ENERGY COLLISIONS}

\vskip 5mm  \underline{V.A. Saleev}$^{1 \dag}$ and D.V. Vasin$^{1}$

\vskip 5mm {\small (1) {\it Samara State University, Samara, Russia
}\\
$\dag$ {\it E-mail: saleev@ssu.samara.ru}}
\end{center}

\vskip 5mm
\begin{abstract}
We consider charmonium polarization at high-energy hadron collider
Tevatron in the framework of the nonrelativistic QCD (NRQCD) and the
$k_T$-factorization approach. The polarization effects are studied
for the direct and the prompt production channels. The obtained
predictions can be used for test of the Regge limit of QCD and for
test of the NRQCD formalism.
\end{abstract}

\vskip 8mm

\centerline{\bf  High-energy factorization}

In the phenomenology of strong interactions at high energy it is
needed to describe the QCD evolution of gluon distribution functions
in colliding particles starting from the scale $\mu_0$, which
controls a nonperturbative regime, to the typical scale of hard
scattering processes $\mu\sim M_T=\sqrt{M^2+|{\bf p}_T|^2}\mbox{,}$
where $M_T$ is the transverse mass of the produced in a hard process
particle. In the region of very high energies
$x=\frac{\mu}{\sqrt{S}}\ll 1\mbox{.}$ This fact leads to the big
logarithmic contributions $\sim (\alpha_s \log(1/x))^n$ in the
resummation procedure, which is described by the BFKL \cite{BFKL}
evolution equation for an unintegrated gluon distribution function
$\Phi(x,|{\bf k}_T|^2,\mu^2)\mbox{.}$

In the $k_T$-factorization approach \cite{KTAproach} hadronic
$\sigma(p + p \to {\cal H} + X)$ and partonic $\hat\sigma(R + R
\to {\cal H} + X)$ cross sections are connected as follows:
\begin{eqnarray}
&&d\sigma(p + p \to {\cal H} + X, S)= \int{\frac{d x_1}{x_1}}
\int{d|{\bf k}_{1T}|^2}\int{\frac{d \varphi_1}{2 \pi}}\Phi(x_1,|{\bf
k}_{1T}|^2,\mu^2) \times\nonumber\\&& \times\int{\frac{d x_2}{x_2}}
\int{d|{\bf k}_{2T}|^2}\int{\frac{d \varphi_2}{2 \pi}}\Phi(x_2,|{\bf
k}_{2T}|^2,\mu^2) d\hat \sigma(R + R \to {\cal H} + X, {\bf
k}_{1T},{\bf k}_{2T})\mbox{,} \label{eq:KT}
\end{eqnarray}
where $x_{1,2}$ are the fractions of the proton momenta passed on to
the gluons, and $\varphi_{1,2}$ are the angles enclosed between
${\bf k}_{1,2T}$ and the transverse momentum ${\bf p}_T$ of ${\cal
H}$, $R$ is considered as {\it reggeized} gluon \cite{KMRK}. In a
stage of numerical calculation we use the following unintegrated
gluon distribution functions $\Phi(x,|{\bf k}_{T}|^2,\mu^2)$: JB
\cite{JB}, JS \cite{JS} and  KMR \cite{KMR}.

The collinear and the unintegrated gluon distribution functions are
formally related as
\begin{equation}
x G(x,\mu^2)\simeq\int{d|{\bf k}_{T}|^2}\Phi(x,|{\bf
k}_{T}|^2,\mu^2)\mbox{.}
\end{equation}
This implies that the cross section (\ref{eq:KT}) is normalized
approximately for the parton model cross section, so that when ${\bf
k}_{1T}={\bf k}_{2T}=0$  we recover the usual gluon-gluon result for
the on-shell gluons.

\centerline{\bf  NRQCD formalism}

In the framework of the NRQCD approach \cite{NRQCD} the heavy
quarkonium ${\cal H}$ production cross section in a partonic
process
 $\hat \sigma (a + b \to {\cal H} + X)$  may be presented as a sum of
terms in which the effects of the long and short distances are
factorized:
\begin{eqnarray}
d\hat \sigma ({\cal H})&=&\sum_n d\hat \sigma (Q\bar Q
[n])\langle{\cal O}^{\cal H}[n]\rangle \mbox{.}
\end{eqnarray}
Here $n$ denotes the set of the color, spin and orbital quantum
numbers of the $Q\bar Q$-pair, $\hat \sigma (Q\bar Q[n])$ is the
cross section of the $Q\bar Q$-pair  production with quantum numbers
$n$ and with the equal 4-momenta. The last one can be calculated
using the perturbative approach of the QCD as an expansion in small
constant of strong interaction $\alpha_s$ and using the
nonrelativistic approximation for the relative motion of the heavy
quarks in the $Q\bar Q$-pair. The nonperturbative transition of the
$Q\bar Q$-pair into the final quarkonium ${\cal H}$ is described by
the long distance matrix element $\langle {\cal O}^{\cal
H}[n]\rangle$. The relevant intermediate states are $[n] =
[{^3S}_1^{(1)}, {^3S}_1^{(8)}, {^1S}_0^{(8)}, {^3P}_J^{(8)}]$ for
${\cal H}
 = J/\psi, \psi'$ and
$[n] = [{^3P}_J^{(1)}, {^3S}_1^{(8)}]$ for  ${\cal H} = \chi_{cJ}$,
where $J=0,1$ and $2$.

\centerline{\bf  Charmonium production by reggeized gluons}

In this part we obtain squared amplitudes for the charmonium
production via the fusion of two reggeized gluons  in the framework
of the NRQCD. We consider the leading order (LO) in $\alpha_s$ and
$v$ contributions of the following  partonic subprocesses:
\begin{equation}
R + R \to {\cal H} [{^3S}_1^{(8)}, {^1S}_0^{(8)}, {^3P}_J^{(1)},
{^3P}_J^{(8)}],\label{eq:RRtoH}
\end{equation}
\begin{equation}
R + R \to {\cal H} [{^3S}_1^{(1)}] + g,\label{eq:RRtoHG}
\end{equation}
The analysis of the next to leading order (NLO) contribution in
the processes of the reggeized gluon-gluon production of the
quarkonia in the $k_T$-factorization approach is outside the
presented paper and it needs a special investigation.

In the case of unpolarized heavy quarkonium production the squared
amplitudes for the different $S$-wave and $P$-wave intermediate
states in the fusion of two reggeized gluons nave been presented
recently in our papers \cite{HSQCD2004}.

\centerline{\bf  Polarization formalism}

In hadronic center-of-mass (CM) reference frame we can define
\cite{KniehlLee, BenekeKramer} the longitudinal polarization
4-vector for spin-one boson (${^3S}_1, {^3P}_1$) by a covariant way:
\begin{equation}
\varepsilon^{\mu}(0)=Z^{\mu}=\frac{(PQ)
P^{\mu}/M-MQ^{\mu}}{\sqrt{(PQ)^2-M^2S}},
\end{equation}
where $Q=P_{1}+P_{2}$, $S=Q^2$, $P_{1,2}$ are colliding hadron
4-momenta. The polarization tensor can be presented as follows:
\begin{eqnarray}
{\cal
P}^{\mu\nu}=\sum_{|\lambda|=0,1}\varepsilon^{\star\mu}(\lambda)\varepsilon^{\nu}(\lambda)=
-g^{\mu\nu}+\frac{P^{\mu}P^{\nu}}{M^2},\\
{\cal P}^{\mu\nu}_0=\varepsilon^{\star\mu}(0)\varepsilon^{\nu}(0)=Z^{\mu}Z^{\nu},\\
{\cal
P}^{\mu\nu}_1=\sum_{|\lambda|=1}^1\varepsilon^{\star\mu}(\lambda)\varepsilon^{\nu}(\lambda)=
{\cal P}^{\mu\nu}-{\cal P}^{\mu\nu}_0
\end{eqnarray}
In the spin-two case ($^3P_2$) the polarization tensor ${\cal
P}^{\mu\nu\rho\sigma}_{|\lambda|}$ reads \cite{ChoWiseTrivedi}:
\begin{eqnarray}
{\cal P}^{\mu\nu\rho\sigma}_0=\frac{1}{6}(2{\cal P}^{\mu\nu}-{\cal
P}_1^{\mu\nu})(2{\cal P}_0^{\rho\sigma}
-{\cal P}_1^{\rho\sigma}),\\
{\cal P}^{\mu\nu\rho\sigma}_1=\frac{1}{2}({\cal P}_0^{\mu\rho}{\cal
P}_1^{\nu\sigma}+ {\cal P}_0^{\mu\sigma}{\cal P}_1^{\nu\rho}+{\cal
P}_0^{\nu\rho}{\cal P}_1^{\mu\sigma}+{\cal P}_0^{\nu\sigma}
{\cal P}_1^{\mu\rho}),\\
{\cal P}^{\mu\nu\rho\sigma}_2=\frac{1}{2}({\cal P}_0^{\mu\rho}{\cal
P}_1^{\nu\sigma}+ {\cal P}_0^{\mu\sigma}{\cal P}_1^{\nu\rho}-{\cal
P}_0^{\mu\nu}{\cal P}_1^{\rho\sigma}).
\end{eqnarray}

As usual, the spin asymmetry parameter is defined as follows
\begin{equation}
\alpha(p_T)=\frac{\sigma_T-2\sigma_L}{\sigma_T+2\sigma_L},
\end{equation}
where
$\sigma_{L,T}=\sigma_{0,1}=\displaystyle{\frac{d\sigma}{dp_T}}(p +
p\to J/\psi_{L,T}X)$ For the polarized $J/\psi$ or $\psi'$
production via direct channel we can write \cite{ChoLeibovich}:
\begin{eqnarray}
\sigma_L^{J/\psi,\psi'}&=&\sigma_0^{J/\psi,\psi'}(^3S_1^{(1)})+\sigma_0^{J/\psi,\psi'}
(^3S_1^{(8)})+\frac{1}{3}\sigma^{J/\psi,\psi'}(^1S_0^{(8)})
+\frac{1}{3}\sigma^{J/\psi,\psi'}(^3P_0^{(8)})+\nonumber\\
&&\frac{1}{2}\sigma_1^{J/\psi,\psi'}(^3P_1^{(8)})
+\frac{2}{3}\sigma_0^{J/\psi,\psi'}(^3P_2^{(8)})+\frac{1}{2}\sigma_1^{J/\psi,\psi'}(^3P_2^{(8)})\nonumber
\end{eqnarray}
As it can be shown \cite{KniehlLee} for the prompt polarized
$J/\psi$ production one has:
\begin{equation}
\sigma_L^{prompt}=\sigma_L^{J/\psi}+\sigma_L^{\chi_c\to
J/\psi}+\sigma_L^{\psi'\to J/\psi}+\sigma_L^{\psi'\to \chi_c\to
J/\psi}\nonumber
\end{equation}
\begin{eqnarray}
&&\sigma_L^{\chi_c\to
J/\psi}=[\frac{1}{3}\sigma^{\chi_{c0}}(^3P_0^{(1)})+\frac{1}{3}\sigma_0^{\chi_{c0}}(^3S_1^{(8)})]Br(\chi_{c0}\to
J/\psi+\gamma)+\nonumber\\
&+&\frac{1}{3}[\frac{1}{2}\sigma_1^{\chi_{c1}}(^3P_1^{(1)})+\frac{1}{2}\sigma_0^{\chi_{c1}}(^3S_1^{(8)})+\frac{1}{4}
\sigma_1^{\chi_{c1}}(^3S_1^{(8)})]Br(\chi_{c1}\to J/\psi+\gamma)+\nonumber\\
&+&\frac{1}{5}[\frac{2}{3}\sigma_0^{\chi_{c2}}(^3P_2^{(1)})+\frac{1}{2}\sigma_1^{\chi_{c2}}(^3P_2^{(1)})+\frac{17}{30}\sigma_0^{\chi_{c2}}(^3S_1^{(8)})+
\frac{13}{60}\sigma_1^{\chi_{c2}}(^3S_1^{(8)})]Br(\chi_{c2}\to
J/\psi+\gamma)\nonumber
\end{eqnarray}
\begin{equation}
\sigma_L^{\psi'\to J/\psi}=\sigma_L^{\psi'} Br(\psi'\to
J/\psi+X)\nonumber
\end{equation}
\begin{eqnarray}
\sigma_L^{\psi'\to\chi_c\to
J/\psi}&=&\frac{1}{3}\sigma_L^{\psi'}Br(\psi'\to
\chi_{c0})Br(\chi_{c0}\to
J/\psi+\gamma)+\nonumber\\
&&+(\frac{1}{2}\sigma_L^{\psi'}+\frac{1}{4}\sigma_T^{\psi'})Br(\psi'\to
\chi_{c1})Br(\chi_{c1}\to
J/\psi+\gamma)+\nonumber\\
&&+(\frac{17}{30}\sigma_L^{\psi'}+\frac{13}{60}\sigma_T^{\psi'})Br(\psi'\to
\chi_{c2})Br(\chi_{c2}\to J/\psi+\gamma)\nonumber
\end{eqnarray}

\begin{figure}[h]
\begin{center}
\begin{tabular}{cc}
\mbox{\epsfig{figure=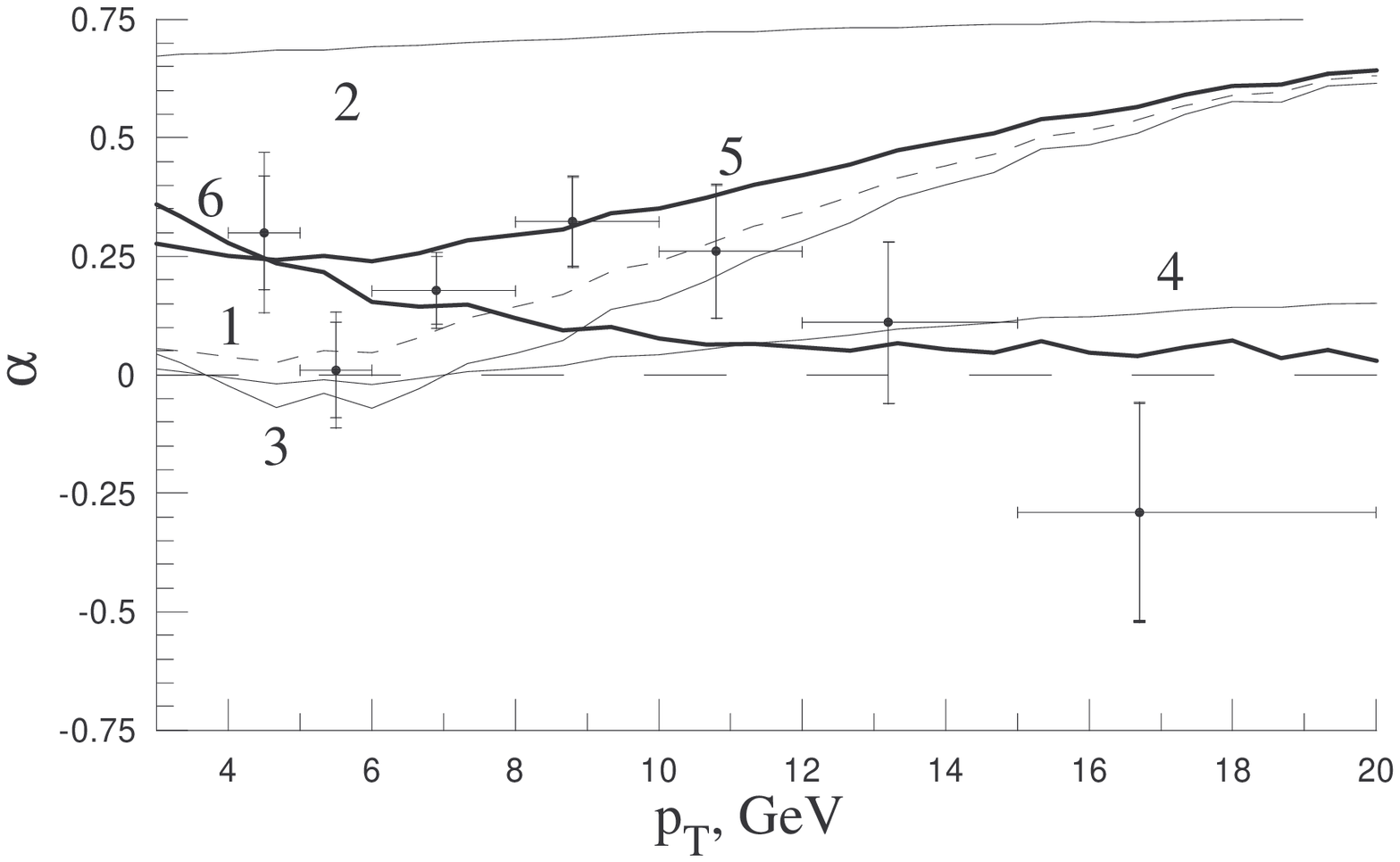,width=0.45\textwidth}}&
\mbox{\epsfig{figure=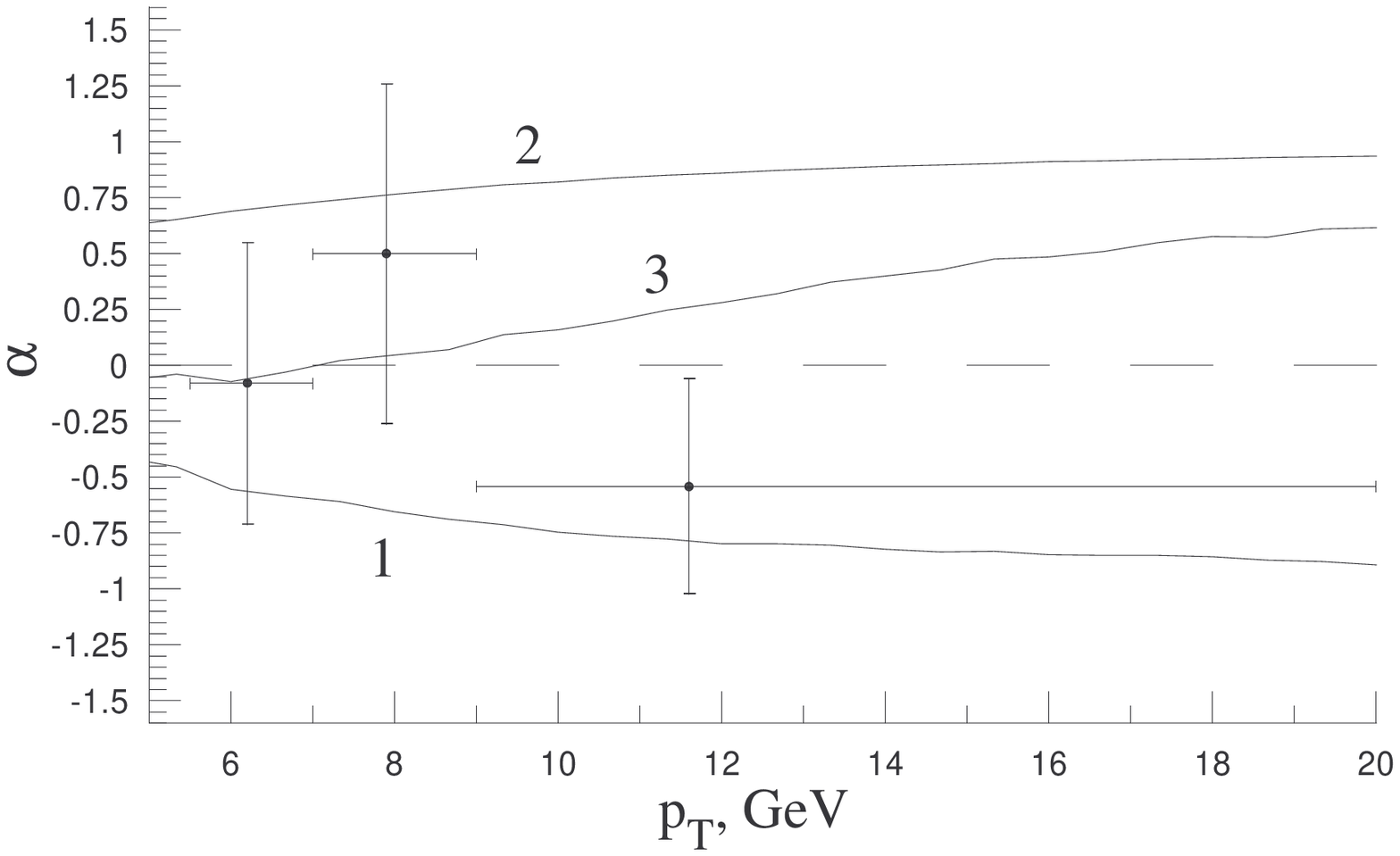,width=0.45\textwidth}}\\
{\bf(a)}& {\bf(b)}
\end{tabular}
\end{center}
{\small{\bf Figure 1a.} Polarization parameter $\alpha(p_T)$ for
prompt $J/\psi$ production. Curve 1 --- the direct production
channel, 2
--- $J/\psi$ from  $\chi_c \to J/\Psi\gamma$ decays,
3 --- $J/\psi$ from $\psi' \to J/\psi$ decays, 4 --- $J/\psi$ from
$\psi' \to \chi_c \to J/\psi$ decays, 5 --- the sum of (1)-(4) terms
, 6 --- the CSM
prediction.}\\
{\small{\bf Figure 1b.} Polarization parameter $\alpha(p_T)$ for
direct $\psi'$ meson production. Curve 1 --- the CSM prediction,
2 --- the color-octet mechanism prediction, 3 --- the direct production channel.}\\
\end{figure}

\centerline{\bf  Charmonium production at Tevatron}

We performed fit Tevatron data \cite{CDFI,CDFII} and obtained sets
of nonperturbative matrix elements (NMEs) (see Table \ref{Tab:NMEs})
\cite{KniehlSaleevVasin}. Using these values we calculated parameter
$\alpha(p_T)$ in cases of different production mechanisms. The
results for  prompt $J/\psi$ and direct $\psi'$ are shown in the
Fig.~1a and Fig.~1b in compare with Tevatron data \cite{CDFSpin}.
Last one was investigated in the collinear parton mode too
\cite{AlphaPM,BKL}. We see that none of these studies were able to
prove or disprove the NRQCD factorization hypothesis.

\begin{table}
 \begin{center}
\caption{NMEs for $J/\psi$, $\psi^\prime$, and $\chi_{cJ}$ mesons
from fits in the collinear parton model (PM) and in the
$k_T$-factiorization approach using the JB \cite{JB}, JS
\cite{JS}, and KMR \cite{KMR} unintegrated gluon distribution
functions. The CDF prompt data from run~I \cite{CDFI} and run~II
\cite{CDFII} have been excluded from our fit based on the JB
gluon density, if these data have been included the
$\chi^2/\mathrm{d.o.f}$ becomes greater then 20.} \vspace{2mm}
 \begin{tabular}{|c|c|c|c|c|}
 \hline
 n $/$ n & PM \cite{BKL} & Fit JB & Fit JS & Fit KMR \\
 \hline
$\langle {\cal O}^{J/\psi}[^3S_1^{(1)}]\rangle/$GeV$^3$ &
1.3 & 1.3 & 1.3 & 1.3 \\
$\langle {\cal O}^{J/\psi}[^3S_1^{(8)}]\rangle/$GeV$^3$ &
$4.4\times10^{-3}$ & $1.5\times10^{-3}$ & $6.1\times10^{-3}$ &
$2.7\times10^{-3}$ \\
$\langle{\cal O}^{J/\psi}[^1S_0^{(8)}]\rangle/$GeV$^3$ &
$4.3\times10^{-2}$ & $6.6\times10^{-3}$ & $9.0\times10^{-3}$ &
$1.4\times10^{-2}$ \\
$\langle {\cal O}^{J/\psi}[^3P_0^{(8)}]\rangle/$GeV$^5$ &
$2.8\times10^{-2}$ & 0 & 0 & 0 \\
\hline $\langle {\cal
O}^{\psi^\prime}[^3S_1^{(1)}]\rangle/$GeV$^3$ &
$6.5\times10^{-1}$ & $6.5\times10^{-1}$ & $6.5\times10^{-1}$ &
$6.5\times10^{-1}$ \\
$\langle {\cal O}^{\psi^\prime}[^3S_1^{(8)}]\rangle/$GeV$^3$ &
$4.2\times10^{-3}$ & $3.0\times10^{-4}$ & $1.5\times10^{-3}$ &
$8.3\times10^{-4}$ \\
$\langle{\cal O}^{\psi^\prime}[^1S_0^{(8)}]\rangle/$GeV$^3$ &
$6.9\times10^{-3}$ & 0 & 0 & 0 \\
$\langle {\cal O}^{\psi^\prime}[^3P_0^{(8)}]\rangle/$GeV$^5$ &
$3.9\times10^{-3}$ & 0 & 0 & 0 \\
\hline $\langle {\cal O}^{\chi_{c0}}[^3P_0^{(1)}]\rangle/$GeV$^5$
& $8.9\times10^{-2}$ & $8.9\times10^{-2}$ & $8.9\times10^{-2}$ &
$8.9\times10^{-2}$ \\
$\langle {\cal O}^{\chi_{c0}}[^3S_1^{(8)}]\rangle/$GeV$^3$ &
$4.4\times10^{-3}$ & 0 & $2.2\times10^{-4}$ & $4.7\times10^{-5}$ \\
\hline
$\chi^2/\mathrm{d.o.f}$ & --  & 2.2 & 4.1 & 3.0 \\
 \hline
 \end{tabular}
 \label{Tab:NMEs}
 \end{center}
  \end{table}

%

\newpage
\centerline{\bf Conclusions} We have obtained analytical formulas
for the squared amplitudes of the processes $R + R\to {\cal H}[1,8]$
and $R + R\to {\cal H}[1]+g$, where ${\cal H}$ may be in the
polarized state. Using new set of the color-octet NMEs  we have
predicted $\alpha (p_T)$ for direct $\psi'$ and prompt $J/\psi$. Our
predictions are coincide with the collinear parton model
calculations rather than with previous $k_T$-factorization results
~\cite{KTspin}.

\centerline{\bf Acknowledgments} We thank  Spin-2005 Organized
Committee for the kind hospitality during the Conference. We also
thank B.~Kniehl, E.~Kuraev and O.~Teryaev for the useful discussion
the questions under consideration in this report.


\begin{thebibliography}{99}
\bibitem{BFKL}
E.~A.~Kuraev, L.~N.~Lipatov, V.~S.~Fadin, Sov.Phys.JETP\
\textbf{44}, 443 (1976) [Zh.Eksp.Teor.Fiz.\  \textbf{71}, 840
(1976)]; I.~I.~Balitsky, L.~N.~Lipatov, Sov.J.Nucl.Phys.\  {\bf
28}, 822 (1978) [Yad.Fiz.\  \textbf{28}, 1597 (1978)].
%
\bibitem{KTAproach}
L.~V.~Gribov, E.~M.~Levin, M.~G.~Ryskin, Phys.Rept.\
\textbf{100}, 1 (1983); J.~C.~Collins, R.~K.~Ellis, Nucl.Phys.\ B
\textbf{360}, 3 (1991); S.~Catani, M.~Ciafoloni, F.~Hautmann,
Nucl.Phys.\ B \textbf{366}, 135 (1991).
%
\bibitem{KMRK}
L.~N.~Lipatov, Nucl.Phys.\  B \textbf{452}, 369 (1995);
E.~N.~Antonov, L.~N.~Lipatov, E.~A.~Kuraev, I.~O.~Cherednikov,
Nucl.Phys.\ B \textbf{721}, 111 (2005).
%
\bibitem{JB}
J.~Bl\"umlein, Report No.\ DESY~95-121 (1995).
%
\bibitem{JS}
H.~Jung, G.~P.~Salam, Eur.Phys.J. C \textbf{19}, 351 (2001).
%
\bibitem{KMR}
M.~A.~Kimber, A.~D.~Martin, M.~G.~Ryskin, Phys.Rev.\  D
\textbf{63}, 114027 (2001).
%
\bibitem{NRQCD}
G.~T.~Bodwin, E.~Braaten, G.~P.~Lepage, Phys.Rev.\ D \textbf{51},
1125 (1995); \textbf{55}, 5853(E) (1997).
%
\bibitem{HSQCD2004}
V.~A.~Saleev, D.~V.~Vasin, Proc. of the First Int. Workshop
HSQCD-2004, 73 (2004).
%
\bibitem{KniehlLee}
B.~A.~Kniehl, J.~Lee, Phys.Rev. D \textbf{62}, 114027 (2000).
%
\bibitem{BenekeKramer}
M.~Beneke, M.~Kr\"amer, M.~V\"anttinen, Phys.Rev. D \textbf{57},
4258 (1998).
%
\bibitem{ChoWiseTrivedi}
P.~Cho, M.~B.~Wise, S.~P.~Trivedi, Phys.Rev. D \textbf{51}, 2039
(1995).
%
\bibitem{ChoLeibovich}
P.~Cho, A.~K.~Leibovich, Report No.\ CALT~68-2026 (1995).
%
\bibitem{CDFI}
CDF Collab., F.~Abe et al., Phys.Rev.Lett.\ \textbf{79}, 572
(1997); \textbf{79}, 578 (1997); CDF Collab., T.~Affolder et al.,
Phys.Rev.Lett.\ \textbf{85}, 2886 (2000).
%
\bibitem{CDFII}
CDF Collab., D.~Acosta et al., Phys.Rev.\ D \textbf{71}, 032001
(2005).
%
\bibitem{KniehlSaleevVasin}
B.~A.~Kniehl, V.~A.~Saleev, D.~V.~Vasin, to be published.
%
\bibitem{CDFSpin}
CDF Collab., T.~Affolder et al., Phys.Rev.Lett.\  \textbf{85},
2886 (2000).

\bibitem{AlphaPM}
M.~Beneke, M.~Kr\"amer, Phys.Rev.\ D \textbf{55}, 5269 (1997);
%
A.~K.~Leibovich, Phys.Rev.\ D \textbf{56}, 4412 (1997).
%
\bibitem{BKL}
E.~Braaten, B.~A.~Kniehl, J.~Lee, Phys.Rev.\ D \textbf{62},
094005 (2000).

\bibitem{KTspin}
 F.~Yuan, K.~T.~Chao, Phys.Rev.\ D \textbf{87}, 022002
(2001).
\end{thebibliography}
\end{document}